\documentclass[aps,pra,twocolumn]{revtex4}
\usepackage{bm}
\usepackage{amsmath}

\usepackage{amssymb}
\usepackage{graphicx}
\usepackage{multirow}

\begin{document}

\newtheorem{theorem}{Theorem}
\newtheorem{corrolary}{Corollary}

\def\pr{\prime}
\def\be{\begin{equation}}
\def\en#1{\label{#1}\end{equation}}
\def\d{\dagger}
\def\bar#1{\overline #1}
\def\S{\mathcal{S}}
\def\D{\mathcal{D}}
\newcommand{\per}{\mathrm{per}}
\newcommand{\rd}{\mathrm{d}}
\newcommand{\vare}{\varepsilon }

\newcommand{\etab}{\bm{\eta}}
\newcommand{\bz}{\mathbf{0}}
\newcommand{\bc}{\hat{\mathbf{c}}}
\newcommand{\m}{\mathbf{m}}
\newcommand{\n}{\mathbf{n}}
\newcommand{\bk}{\mathbf{k}}
\newcommand{\bl}{\mathbf{l}}
\newcommand{\petal}{\frac{\partial }{\partial \eta_l}}
\newcommand{\alphb}{\bm{\alpha}}
\newcommand{\V}{{\mathcal{V}}}

\title{Collective phases  of identical particles  interfering on    linear multiports}

\author{V. S. Shchesnovich and M. E. O. Bezerra}

\affiliation{Centro de Ci\^encias Naturais e Humanas, Universidade Federal do
ABC, Santo Andr\'e,  SP, 09210-170 Brazil }

\begin{abstract}
We introduce collective geometric   phases of bosons and fermions  interfering  on a  linear unitary multiport, where each phase depends on the internal states of identical particles (i.e.,  not affected by the multiport)  and corresponds to  a cycle  of the symmetric group.   We  show   that  quantum interference of $N$ particles in  generic pure internal states, i.e., with no pair being  orthogonal,   is governed by $(N-1)(N-2)/2$ independent  triad phases (each involving only three particles).  The deterministic distinguishability,  preventing  quantum interference with two or three particles,  allows for the genuine $(N\ge 4)$-particle phase (interference) on a multiport:  setting   each particle to be deterministically distinguishable from all others except two  by their internal states allows for   a novel (circle-dance) interference of $N\ge 4$ particles governed by a collective $N$-particle phase, while simultaneously preventing   the $R$-particle interference for $3\le R\le N-1$. The   genuine  $N$-particle interference  manifests the  $N$th order quantum correlations between identical particles at a multiport output, it  does not appear in the marginal probability for a subset  of the particles, e.g., it  cannot be detected  if at least  one of the particles is lost. This means that  the   collective phases are  not detectable by the usual   ``quantumness"  criteria based on the  second-order quantum correlations.  The results can be useful  for   quantum computation, quantum information, and other quantum technologies  with single photons.  \end{abstract}

\maketitle

\section{Introduction}
The Hong-Ou-Mandel  experiment \cite{HOM} with two single  photons,   recently repeated with neutral atoms \cite{AHOM}, manifests   the proportionality relation  between the visibility of    interference    and the degree of partial distinguishability due to  internal states of bosons  \cite{Mandel1991}.     To extend   this fundamental relation for   $N$ identical bosons and  fermions \cite{FHOM} is  an important  fundamental problem with  applications in the fields of  quantum  information and computation, quantum    state engineering, and quantum metrology \cite{KLM,Peruzzo,AA,E1,E2,E3,E4,Metcalf,ULO,RWPh,QMet}.
The  theory has advanced considerably in recent years  \cite{MPBF,SUN,PartDist,Tichy,Tamma,Rohde,MCMS,BB,GL,CrSp},  however our  understanding of the  relation between partial distinguishability of $N$ identical particles and their   interference on a multiport is still not complete.  Non-trivial  quantum-to-classical transition of  more than two photons \cite{4PhDist,4PhExp,NonM} and  the   recent observation of a collective (triad) geometric phase in the genuine three-photon interference   \cite{Triad2}, i.e.,  a phase attributed to the three photons as a whole,   add to the complexity of the problem. The triad  phase can be understood as a multi-particle realization of  the Pancharatnam-Berry phase  \cite{Panch,Berry} in the internal Hilbert space, it is also defined quite similarly to  the Bargmann geometric invariant \cite{Bargm,ExpBargm} for three quantum states.    Therefore,  distinguishability of identical particles  is a global property that cannot be reduced to considering distinguishability of only pairs of states \cite{JS}. There is also similarity to the   fully entangled  $N$-particle state, which   exhibits the  genuine $N$-particle interference with  a  collective phase \cite{GHSZ,GHZ,MInt} (due to phases in the individual  Mach-Zehnder  interferometers), demonstrated in another recent  experiment  \cite{Triad1}. In view of these important  relations, the collective  phases   of identical particles deserve to be thoroughly investigated. Is it possible to have   multiparticle   collective phases for  $N$ identical particles,   independent of the collective phases of $R<N$ particles,  and how to arrange for such a case?   How to approach the characterization of the multiparticle interference and collective phases  in  the general   case? Can collective phases be detected by popular  criteria of quantum behavior? We  formulate a  general framework able to  provide  answers to the above  questions and  explore the relation  of distinguishability due to  the  internal state discrimination    \cite{Hel,Chefles} of identical particles and their interference on a multiport.   We also find that weighted  graph theory illustrates    the relation between  partial distinguishability and multiparticle   interference and allows to simplify the proofs of   some of the presented results. 

In section \ref{sec2} we discuss  the general framework for our analysis, introduce the notion of multiparticle interference, define what we call the genuine $N$-particle interference for $N\ge 3$,  and make a connection between  the weighted graph theory (in a generalized form) and  the  partial distinguishability of independent identical particles in general (mixed) internal states. In section \ref{sec3} we concentrate on  identical particles in pure internal states, introduce the notion of a collective multiparticle phase,  prove two theorems  on the existence of the  genuine (circle-dance) multiparticle interference, where we analyze  in detail the case of $N=4$ and give   an example of the circle-dance interference with single photons in Gaussian spectral states. In section \ref{sec4} we show that  the collective $N$-particle  phase governing the circle-dance interference  of identical particles  is a  consequence of the  genuine $N$th-order quantum correlations between them.  Section \ref{sec5} contains the conclusion. Some mathematical details from sections \ref{sec2}-\ref{sec4}  are relegated to  the appendices.

 \section{Permutation cycles, multiparticle interference and graph theory}
 \label{sec2}

We consider  interference on a unitary multiport of   noninteracting  identical particles  (either bosons or fermions) coming from independent sources. E.g., in the case of single photons,  a multiport can be a spatial arrangement of beam splitters and phase shifters, or an integrated optical network,  as in Refs. \cite{KLM,E1,E2,E3,E4,Metcalf,ULO,RWPh}.    Applications are possible  even in the case of interacting particles, e.g., to the  multiparticle scattering   in a  fixed number of discrete  channels \cite{MCMS},   where the    scattering  into a set of   discrete channels plays the role of a  unitary transformation.   

 We fix the number of input and output ports of a multiport to be  $M$ and the total number of particles to be $N$. We assume that the input to a multiport is given  by the  states $|k_1^{(a)}\rangle \langle k_1^{(a)}| \otimes \rho^{(1)}, \ldots, |k_N^{(a)}\rangle  \langle k_N^{(a)}| \otimes\rho^{(N)}$, where $\rho^{(i)}$   is the  internal state of particle $i$   and $|k_i^{(a)}\rangle$ stands for the quantum  mode of input port $k_i$ of the multiport.  A multiport   performs   a unitary transformation ($U$) between the input   and output  modes as follows   $|k^{(a)}\rangle  = \sum_{l=1}^M U_{kl}|l^{(b)}\rangle$,  or in the  second-quantization notation   $\hat{a}^\dag_{k,j} = \sum_{l=1}^MU_{kl}\hat{b}^\dag_{l,j}$, where $\hat{a}^\dag_{k,j}$ ($\hat{b}^\dag_{l,j}$) creates a particle in the input  mode $|k^{(a)}\rangle$  (respectively, output mode $|l^{(b)}\rangle$) and an internal state $|j\rangle\in \mathcal{H}$.

Throughout the text,  when discussing the input and output of a multiport, we will  use  the following notations:  the vector $\bk = (k_1,\ldots,k_N)$ will stand for the sequence of input ports occupied by particles (arranged in  nondecreasing order, when there are more than one particle per port), whereas the vector $\bl = (l_1,\ldots,l_N)$ will stand for the same for the output ports. 
 We will also use   $\n = (n_1,\ldots,n_M)$ and $\m = (m_1,\ldots,m_M)$ for the occupations of the input and output ports, respectively, where $n_i$  ($m_i$) denote the number of particles in the input (output) port $i$.  In the main text only  the  input  with up to one particle per port is considered, with  the input ports fixed  to be  $\bk=(1,\ldots,N)$, fig.~\ref{F1}(a) (arbitrary  input configuration $\n = (n_1,\ldots,n_M)$  is  considered  in appendix \ref{appA}).

 \begin{figure}[htb]
\begin{center}
\includegraphics[width=0.5\textwidth]{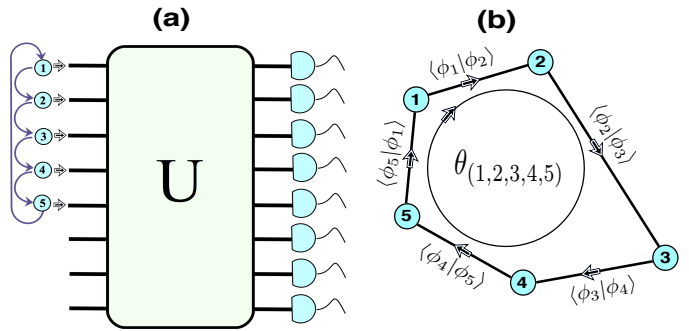}
\caption{\textbf{(a)}  $N$   identical particles are impinging on a multiport with a unitary matrix $U$.  The circular arrows illustrate the  $N$-particle cycle $(1,2,3,\ldots,N)$  responsible  for the circle-dance  interference  with  deterministically distinguishable particles $\alpha$ and $\beta$ for $\beta\ne \alpha\pm 1$ (\textit{mod} $N$).   \textbf{(b)}  A weighted directed graph representation of particle distinguishability  in panel \textbf{(a)}  in the  case of  pure internal states $|\phi_k\rangle$, $k=1,\ldots,N$, where each  particle is a vertex and    a directed edge $k\to l$ has the complex weight   $w(k,l)\equiv - \ln(\langle\phi_l|\phi_{k}\rangle) = d_{kl} + i\theta_{kl}$, where $d_{kl}$ serves as the distance between the vertices (here $d_{kl} =\infty$ is indicated by the absence of such an  edge) and $\theta_{kl}$  is the phase in  the direction $k\to l$.    Only   two   $(R\ge 3)$-particle    cycles  on a finite path --  $1\to2\to \ldots \to 5 \to 1$ and its inverse -- contribute to output probability in panel \textbf{(a)}, therefore the latter  depends on a single  collective (five-particle) phase along the edges  $\theta_{(1,2,3,4,5)} =  \theta_{12} +  \theta_{23} +  \theta_{34} +  \theta_{45} + \theta_{51}$  (see the text).}\label{F1}
\end{center}
\end{figure}

 The internal states  define  identical particle distinguishability which  affects their interference  on a linear multiport. For example, in the HOM experiment \cite{HOM} the main source of photon distinguishability was the arrival time which was  recorded   to relate it to the dip in the coincidence counting. One can   characterize the  state of partial distinguishability of identical particles by  the  degree of   possible internal state  discrimination.   We are mainly  interested in the probability of an output configuration $\m$  without account of  the internal states  (i.e.,  when the internal states are not resolved by the particle counting  detectors or simply ignored) which is mathematically expressed by summation over the probabilities with resolved internal states~\footnote{For example, in the HOM  experiment  case we would be  interested in  the probability of two single photons   leaving  certain output ports $l_1$ and $l_2$ of a balanced beam splitter, and not in the probability of one photon leaving mode $l_1$ at time $\tau_1$ and  the other  leaving port $l_2$ at time $\tau_2$.}.   The relation between discrimination of the internal states and its effect on multiparticle  interference  will be briefly considered   in section \ref{sec4} (particle counting with discrimination of the  internal states  is also analyzed to the necessary detail  in appendix \ref{appA}).  Single photons on an optical multiport, as in Refs. \cite{KLM,E1,E2,E3,E4,Metcalf,ULO,RWPh}, are the main application of the theory.  To some extent, the internal state discrimination is  routinely done    with single photons  (e.g., detection of the   time of arrival serves as the   partial state discrimination). 
 
  We  will use the fact  \cite{PartDist} that  just one  complex-valued function on the symmetric group $ \S_N$ for $N$ particles accounts  for  the effect of partial distinguishability, if one  ignores the information on the internal states at a multiport output (this function can be though of as a generalization of Mandel's interference  visibility $\&$  distinguishability parameter  \cite{HOM,Mandel1991} for $N\ge 3$ photons; for full discussion the reader should consult Ref. \cite{PartDist}).  The exact form of  probability of an output configuration $\m$ for interfering identical particles on a linear multiport   was studied before   \cite{PartDist,BB} (for  completeness,   we also provide brief derivation in appendix \ref{appA}). For   independent  particles in internal states $\rho^{(1)},\ldots, \rho^{(N)}$ impinging on the input ports $\bk = (1,\ldots,N)$ of a unitary multiport $U$,  the probability to detect a configuration $\m= (m_1,\ldots,m_M)$   in the  output ports  reads \cite{PartDist,BB}
\be
p_N(\bl|\bk)  = \frac{1}{\m!} \sum_{\tau,\sigma\in\S_N}J(\tau^{-1}\sigma)
  \prod_{k=1}^N U^*_{k,l_{\tau(k)}}U_{k,l_{\sigma(k)}},
\en{E1}
where $\m\equiv m_1!\ldots m_M!$  and the $J$-function accounts for particle  distinguishability.   In our case it   factorizes \cite{PartDist} according to the disjoint cycle decomposition (see, for instance, Ref. \cite{Stanley}) of its argument.  Denoting by $cyc(\sigma)$  the set of  disjoint cycles of a permutation~$\sigma$, we get
\begin{eqnarray}
\label{E3}
J(\sigma) &=& \prod_{\nu\in cyc(\sigma)} g(\nu),\nonumber\\
g(\nu)& \equiv& (\pm1)^{|\nu|-1} \mathrm{Tr}\bigl( \rho^{(k_{|\nu|})}\rho^{(k_{|\nu|-1})}\ldots \rho^{(k_{1})}\bigr),
\end{eqnarray}
 where $\nu= (k_1,\ldots,k_{|\nu|})$, $k_\alpha\in \{1,\ldots,N\}$, stands for    the cycle $k_1\to k_2\to \ldots \to k_{|\nu|}\to k_1$ and   $|\nu|$  for its  length, the  minus sign is due to the signature of a cycle  $\mathrm{sgn}(\nu) = (-1)^{|\nu|-1}$  \cite{Stanley} and applies to fermions.

The specific form of probability in    Eqs. (\ref{E1}) and (\ref{E3}) can be understood without a detailed derivation.    Indeed, for identical   particles  the probability $p_N(\bl|\bk)$   must be  symmetric under  permutations of either  output  ports $\bl$ or  input ports $\bk$ (particles do not  have labels).  Let us  consider  a specific    $N$-particle  transition amplitude on a multiport $\mathcal{A}(\bk\to \bl)\equiv \prod_{k=1}^N U_{k,l_{k}} $  with $\bl = (l_1,\ldots,l_N)$   (i.e., particle $k$ goes to output port $l_k$). By the above  symmetry of  probability, any  permutation $\sigma$  of identical particles  over the output  ports gives another valid  transition amplitude contributing to the probability,  $\mathcal{A}({\bk\to\sigma(\bl)})= \prod_{k=1}^N U_{k,l_{\sigma(k)}}$.  The probability, linear in the amplitude and its complex conjugate,  is given by  the  summation  over two  permutations $\sigma, \tau\in \S_N$ in the two amplitudes as in  Eq. (\ref{E1}) (with the signature of a permutation in the case of fermions), where there is also a   factor equal to  the scalar product in $\mathcal{H}^{\otimes N}$ of the internal states, similarly permuted, with the result  dependent only on the relative permutation,  described by the   function $J(\tau^{-1}\sigma)$ in Eq. (\ref{E1}).  The disjoint cycles of $\tau^{-1}\sigma$  contribute independent factors, since  they permute different particles (the cross-cycle particle distinguishability does not contribute), hence    the $J$-function must be in the form of  Eq. (\ref{E3}) (which is easily established by considering  pure internal states).     Finally, by using the arbitrary permutations, we have permuted $m_l$  particles in output port $l$ as well,  thus  have performed the multiple  counting of  identical terms,  hence the factor $\m!$ in the denominator. 

Due to  the  mutual  independence of the concept of  distinguishability of identical particles and the unitary transformation employed by a multiport, below we will focus on the  state of particle distinguishability  itself when discussing multiparticle interference on a multiport. Indeed, on a  generic multiport, i.e.,  when  none of the matrix elements $U_{kl}$ is zero,  all permutations of particles  can  contribute to the coincidence count output  probability and one can observe the discussed examples of multiparticle interference on such a multiport (obviously, an optimization is possible by selecting a particular multiport).  We will return to this when discussing a specific example in section \ref{sec3}.

From the above discussion, one can derive the physical meaning of the  cycles in Eq. (\ref{E3}):   to each such  cycle    can be associated the  interference  of \textit{only} the particles involved in  the cycle. Below  we will frequently use the term ``$R$-particle interference"  which simply  means the contribution from  an  $R$-cycle in Eq. (\ref{E3})  to  the output probability in Eq. (\ref{E1}). Note, however, that a general summation term in Eq.  (\ref{E3})  consists of simultaneous and independent interferences of $S_1,\ldots,S_d$ particles according to a partition of $N = S_1+\ldots +S_d$.

The relative significance  of the contribution of  $R$-particle interference depends on the $g$-weights of the $R$-cycles in the respective probability.  Obviously, there is always two-particle interference, unless the particles  have orthogonal internal states, i.e.,  $\rho^{(k)}\rho^{(l)}= 0$ for  all $k\ne l$ (distinguishable particles, i.e., the classical  case \cite{PartDist}). It may happen that some  cycles have zero $g$-weight  such that there is just two-particle and $N$-particle interference on a multiport. We will say that this case realizes the ``genuine $N$-particle interference", meaning that no $3\le R\le N-1$ interference is realized at the same time.

Obviously, no   $N$-particle interference  contributes to the output  probabilities, if one sends a subset of $N$ particles to a multiport.  Similarly,  no $N$-particle interference contributes  to the respective (marginal) output probabilities,  when   $N$ particles  are sent to a multiport input, but  at  the output the information on  some of the  particles is lost, or discarded (by binning together  the output configurations $\m$ containing a given configuration  $\m^\prime $ of  less than $N$ particles). This is due to the simple fact that the marginal probability of an  output configuration  for $R$ out of $N$ particles depends \textit{only} on the $d$-cycles with $d\le R$. Indeed, one can use the above discussion of Eqs. (\ref{E1})-(\ref{E3}) to arrive at this conclusion. The probability $p_N(\bl^{\prime}|\bk)$ of an   output $\m^\prime$ with  $R<N$ particles $\bl^\prime=(l^\prime_1,\ldots,l^\prime_R)$ is  obtained by partitioning the transition $\bk\to \bl$ as $(\bk^\prime\to\bl^\prime,\bk^{\prime\prime}\to\bl^{\prime\prime})$, summing up over   $\bl^{\prime\prime}$ (which simply removes the respective matrix elements $U_{kl^{\prime\prime}}$ by the unitarity of $U$) and  averaging  over all $(R,N-R)$-partitions  $(\bk^{\prime},\bk^{\prime\prime})$ of the input ports, i.e., 
\be
p_N(\bl^{\prime}|\bk) 
 = \left({N \atop R}\right)^{-1} \sum_{ \bk^{\prime}\subset \bk } p_R(\bl^{\prime}|\bk^{\prime}), 
\en{E10}
 where the summation is over all subsets  $\bk^\prime$  having  $R$ indices    (for a formal mathematical derivation, see  appendix \ref{appD}). In Eq. (\ref{E10})  $p_R(\bl^{\prime}|\bk^{\prime})$ is the probability of $R$ particles at  input ports $\bk^{\prime}$  to end up at  output ports  $\bl^{\prime\prime}$, given similarly as in  Eqs. (\ref{E1}) and  (\ref{E3}) but now  for $R$ particles, thus it depends only on $d$-cycles with $d\le R$.

By using the Cauchy-Schwartz inequality for the trace-product of operators, one can prove  an important upper bound on the $R$-cycle $g$-weight   by the 2-cycle $g$-weights with the same particles  (see appendix \ref{appB})
\be
 |\mathrm{Tr}(\rho^{(k_1)}\ldots \rho^{(k_R)})|^2 \le \prod_{\alpha=1}^R \mathrm{Tr}(\rho^{(k_\alpha)}\rho^{(k_{\alpha+1})})
\en{E11}
where $\alpha$ is   $\textit{mod}\; R$. 
Up to now, the treatment of identical particle distinguishability was based mainly on combinatorics (permutations), but Eq. (\ref{E11}) suggests   a graph interpretation.  This, however,  requires generalizing the concept of a graph.   Let us think of identical  particles as vertices, with vertex $i$ being  associated with  internal state $\rho^{(i)}$ and all the vertices being connected by edges.  Our main object of study is a cycle    $\nu= (k_1,\ldots, k_R)$  on the edges connecting  vertices  $k_1,\ldots,k_R$ to which we set  a complex weight   
\begin{eqnarray}
\label{weight}
w(\nu) &\equiv& -\ln\left(\mathrm{Tr}\bigl\{ \rho^{(k_{|\nu|})}\rho^{(k_{|\nu|-1})}\ldots \rho^{(k_{1})}\bigr\}\right) \nonumber\\
& = & D_{(k_1,\ldots,k_R)} + i\theta_{(k_1,\ldots,k_R)}, 
\end{eqnarray}
 where $D_{(k_1,\ldots,k_R)}$ is the path length of the cycle, whereas $\theta_{(k_1,\ldots,k_R)}$  (see also fig. \ref{F1}(b)) we will call   the collective $R$-particle phase  of the cycle. Note that reversing the cycle  orientation  changes the sign of  the  cycle phase (a nonzero cycle  phase selects a direction of the path along the cycle).   
 Larger path distance of a cycle means smaller  contribution to output probability on a multiport 
 via Eq.~(\ref{E3}), whereas   Eq. (\ref{E11})   bounds twice the  path distance of a higher order cycle   by    two-vertex cycles on the same edges  (a generalization of  the  usual path addition on a graph).  In the case of pure internal states, when the inequality in Eq. (\ref{E11}) becomes equality, one recovers the  usual additive  distance on a  weighted   graph  and the phases lead to an additional directed graph (see the next section). 
 
The completely indistinguishable  particles and  the distinguishable  (classical) particles     have degenerated graphs.  The   completely indistinguishable particles from independent sources have the same pure internal state   (see Refs. \cite{PartDist,BB})  and map   to the zero-distance graph with all the vertices coinciding (which makes sense, since the particles are completely indistinguishable).    The  deterministically distinguishable particles, $\rho^{(i)}\rho^{(j)}=0$,  for $i\ne j$,  are  mapped to  a set of   vertices   lying at infinite distance  from each other (the infinite distance between two vertices   will be illustrated  as absence of the respective edge, as in fig. \ref{F1}(b)).

\section{Particles in pure internal states  and   $N$-particle interference}
\label{sec3}

Consider particles  in pure internal states $\rho^{(k)} = |\phi_k\rangle\langle \phi_k|$, $k=1,\ldots,N$.    Let us set $ \langle \phi_k|\phi_l\rangle \equiv r_{kl}e^{i\theta_{kl}} = \exp\{-d_{kl} +i \theta_{kl}\}$, where $r_{kl} = e^{-d_{kl}}$ is the absolute value and $\theta_{kl} = -\theta_{lk}$  is the phase of the inner product. We    will  call $\theta_{kl}$  the  mutual  phase of particle $l$ with respect to $k$.   For  the  path  distance along   an $R$-particle cycle  $(k_1,\ldots,k_R)$ we now have from Eq. (\ref{weight})
\be
D_{(k_1,\ldots,k_R)} =\sum_{\alpha=1}^R d_{k_\alpha,k_{\alpha+1}},
\en{gGraph}
whereas for the   collective  $R$-particle phase along the cycle   
\be 
\theta_{(k_1,\ldots,k_R) } = \theta_{k_1 k_2} + \theta_{k_2 k_3} + \ldots + \theta_{k_R k_1}
\en{Rphase}
(note the difference between the two concepts of the phase: for  $R=2$ the mutual phase $\theta_{kl}$ can be arbitrary, whereas   $\theta_{(k,l)}=0$).

\begin{figure}[htb]
\begin{center}
\includegraphics[width=0.35\textwidth]{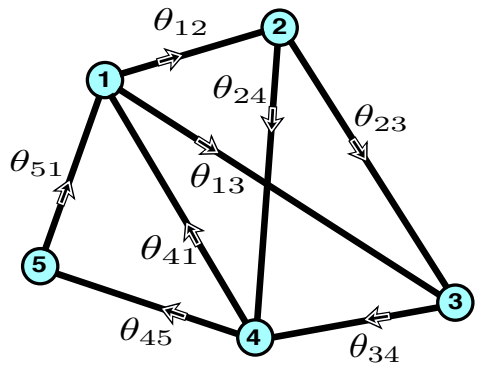}
\caption{The directed   graph  for the  multiparticle phases,   where the  mutual phase $\theta_{kl} \equiv \mathrm{arg}(\langle\phi_k|\phi_{l}\rangle)$ weights the directed edge $k\to l$.   An $(R\ge 3)$-particle  collective phase corresponds to a  closed oriented path on $R$ vertices. Each closed path on edges has a corresponding multiparticle phase obtained by summation of the mutual phases on the edges passed  (with the minus sign, when passing an edge in the inverse direction to  the indicated one). The directed   paths  $1\to 2 \to 3\to 1$, $1\to3\to4\to 1$ and $1\to 2\to4\to 1$   allow to express  any closed  oriented  paths on  the edges  $1,2,3,4$.   }\label{F2}
\end{center}
\end{figure}

We can interpret $d_{kl}$  as the distance between the vertices $k$ and $l$ in a usual distance-weighted graph, whereas the collective phase $\theta_{(k_1,\ldots,k_R)}$ (\ref{Rphase})    leads to a weighted  directed graph, see fig.~\ref{F2}, where each directed edge $k\to l$ acquires a real weight  $\theta_{kl}$.   The   two-graph representation is  just a single directed graph with  complex weights on the   edges given by a Hermitian  adjacency matrix $w_{kl}\equiv -\ln(\langle \phi_l|\phi_k\rangle) = d_{kl} + i\theta_{kl}$. 

Consider  the directed  graph representation of the collective phases, assuming that none of $ r_{kl} $ is zero, i.e., the internal states are only partially distinguishable by the state discrimination  \cite{Hel,Chefles}.    The two-particle collective  phase  $\theta_{(k,l)}  = 0$, since the two directed edges   $k\to l$ and $l\to k$ cancel each other.  For $N=3$ all permutations in $\S_3$ are 
cycles themselves. Since the  collective phases of three particleson the same vertices can differ only by a sign (transposition of two particles reverses the path orientation in fig.~\ref{F2}, or  $\theta_{(i,j,k)} = \mathrm{sgn}(\sigma)\theta_{(1,2,3)}$, for $\sigma(1,2,3) = (i,j,k)$), consider the 3-cycle $\nu=(1,2,3)$.  It has   the   phase $\theta_{(1,2,3)}  \equiv \theta_{12} + \theta_{23} +\theta_{31}$. From the above discussion and section \ref{sec2}  it is now clear that precisely this  phase governs   the genuine three-particle interference of Ref. \cite{Triad2}, observed experimentally    as a variation in the output probability  according to this  phase (by keeping $r_{kl}$ fixed, while varying the mutual  phases $\theta_{kl}$). 

We are interested in the  setups  when there is a  collective $N$-particle phase, which  cannot be  reconstructed with  detection of  less   than $N$ particles, e.g.,  from any marginal probability. In such a case, we will   call such a phase the ``genuine $N$-particle phase". This notion is  similar  to  the triad phase  introduced in   Ref. \cite{Triad2}.     From the above discussion and section \ref{sec2}, it follows  that the two notions coincide for $N=3$. Moreover,  as a  genuine  $(N\ge 4)$-particle phase  can serve only the   $N$-particle collective phase
$\theta_{(k_1,\ldots,k_N)}$, when it is independent of all the $R$-particle collective phases $\theta_{(l_1,\ldots,l_R)}$  for $3\le R\le N-1$.   We can now  state our first result.

\medskip
\noindent\textit{Theorem 1.--} Identical particles  in pure internal states with no two states being orthogonal do not allow for  a genuine  $N$-particle phase  in an interference experiment on a  linear  multiport.

\textit{Proof.--} By the above discussion, the  theorem  will follow from analysis of  the collective phases  in  Eq.~(\ref{Rphase}).   By the linear relations between the mutual and collective phases, the set of all  cycles contains   as many independent collective phases as there are independent  mutual  phases $\theta_{kl}$, i.e., exactly \mbox{$(N-1)(N-2)/2$} phases. Indeed, since the  mutual phases $\theta_{kl}$ are defined up to the global   phase of a state, exactly \mbox{$N-1$} of them can be preset to given values  by employing   the   global phase  transformation $|\phi_k\rangle \to e^{-i\theta_k}|\phi_k\rangle$ resulting in  a phase shift   $\theta_{kl} \to \theta_{kl}+\theta_k - \theta_l$. We can  now state the following lemma, which implies theorem 1.

\textit{Lemma 1.--} For $N$ particles in non-orthogonal pure internal states there are  $(N-1)(N-2)/2$ independent triad phases, e.g., $\theta_{(1,k,l)}$ for $2\le k< l \le N$.  Therefore, all  $R$-particle phases,  $3\le R \le N$, are  linear combinations of the triad phases.

\textit{Proof.--}  Let us start with  $N=4$ particles, which corresponds to the  fully-connected    subgraph on the vertices  $1,2,3,4$  in fig \ref{F2}.  In this case, we have eight  different $3$-cycles, i.e., closed oriented paths on three edges, divided into  two groups of four  phases,  $\theta_{(1,2,3)}$, $\theta_{(1,3,4)}$, $\theta_{(1,2,4)}$, and $\theta_{(2,3,4)}$ and the sign-inverted of these (obtained by  transposition of two particles in a cycle).  Moreover, we have 
\be
\theta_{(2,3,4)} = \theta_{(1,2,3)} + \theta_{(1,3,4)} - \theta_{(1,2,4)}.
\en{E6}
We need to show that we can express  three  independent mutual  phases   as functions of the triad phases $\theta_{(1,2,3)}$,  $\theta_{(1,3,4)}$ and $\theta_{(1,2,4)}$. Selecting the  mutual phases $\theta_{12}$, $\theta_{23}$, and $\theta_{34}$ as the basis (while setting the rest mutual phases  to zero, by using the arbitrariness of the global  phase  of an internal state) we obtain 
\be
\theta_{12} = \theta_{(1,2,4)},\; \theta_{23} = \theta_{(1,2,3)}-\theta_{(1,2,4)},\; \theta_{34} = \theta_{(1,3,4)}.
\en{Ph2}
Thus  all $R$-particle phases are expressed through the phases $\theta_{(1,2,3)}$,  $\theta_{(1,3,4)}$ and $\theta_{(1,2,4)}$, which concludes the proof for $N=4$.

 Consider now $N> 4$ particles (it is enough to consider just five particles, as   in  fig. \ref{F2}).  We have to show that the set of phases $\theta_{(1,k,l)}$ is the basis for all possible triad phases.  For any   triad phase  $\theta_{(i,k,l)}$,  in the graph representation the corresponding oriented path     lies within   the  complete  subgraph  on the  vertices $1,i,k,l$, which returns us to $N\le 4$ particles $1,i,k,l$, i.e., to the case considered above.  Q.E.D.

In section \ref{sec2} we have introduced the notion of the  genuine $N$-particle interference and in this section the notion of the genuine $N$-particle phase. Theorem 1   relates the two  notions for pure internal states, by stating, in other words,  that the genuine $N$-particle phase could be found    only in a genuine $N$-particle interference, i.e., when the $R$-particle interferences are forbidden  for all $3\le R\le N-1$. (We do not know if this relation extends  to  \textit{mixed} internal states, since  the collective phase of  a  higher-order cycle in  Eq.  (\ref{weight}) has no simple linear dependence on the collective phases of  lower-order  cycles on the same edges.)

\subsection{Genuine $N$-particle phase and distinguishability}   
\label{sec3A}

Theorem  1 states that a genuine $(N\ge 4)$-particles phase   may appear only if  some particles have orthogonal internal states, i.e., when they behave  as distinguishable   classical particles with respect to each other, since their  internal states   can  be deterministically discriminated.  The   deterministic distinguishability  prevents  quantum interference with two  particles \cite{HOM,Mandel1991}. For $N=3$,  removing one edge prevents  existence of the $3$-cycles and therefore a three-particle interference. On the other hand, the genuine $(N\ge 4)$-particle interference is possible   when the respective graph consists of just one cycle,  see fig. \ref{F1}(b), which  would prevent    $R$-particle interference for $3\le R \le N-1$ simply by the absence of an edge.  Such an interference   would differ  conceptually   from both the three-particle interference observed in Ref.  \cite{Triad2}  and  from the  genuine $N$-particle interference from  the fully-entangled $N$-particle state \cite{GHSZ,MInt} reported in Ref.~\cite{Triad1} with three photons (by requiring the   deterministic distinguishability of particles).

We must find  conditions on the internal states of particles that allow for a variable $N$-particle phase (\ref{Rphase}) while the corresponding  graph consists of a single cycle.    
A given  set of parameters $\{r_{kl}\ge 0, 0\le \theta_{kl}< 2\pi; 1\le k<l\le N\}$ comes from the inner products of $N$ vectors $|\phi_k\rangle$, $k=1,\ldots, N$,  i.e., satisfies  $\langle\phi_k|\phi_l\rangle  = r_{kl}e^{i\theta_{kl}}$,    if and only if the   respective matrix  $H_{kl}\equiv r_{kl}e^{i\theta_{kl}}$ (with $H_{kk}=1$)  is positive semidefinite  Hermitian  matrix.  The set of conditions $\sum_{l=1 }^N(1-\delta_{kl})  |H_{kl}| =\sum_{l=1}^N (1-\delta_{kl})r_{kl} \le 1 $  for $k=1,\ldots,N$  (here $\delta_{kl}$ is the Kronecker delta) is   sufficient   for $H$  to be such a  matrix, since the   eigenvalues $\lambda_1,\ldots,\lambda_N$ of $H$ are bounded as follows  (known as the Gershgorin  circle theorem \cite{Gersh}) 
\be
|\lambda_k - 1| \le \sum_{l=1}^N (1-\delta_{kl})r_{kl} \le 1
\en{EQ3}
 and therefore are all non-negative. Note that the mutual phases $\theta_{kl}$ remain free parameters.  We can now formulate our second result. 
 
 \medskip
\noindent\textit{Theorem 2.--} Identical particles in  linearly independent internal states    $|\phi_1\rangle, \ldots, |\phi_N\rangle$, where each state being orthogonal to  all others except two, 
\begin{eqnarray}
\label{E9}
 \langle\phi_k|\phi_{k\pm 1}\rangle \ne 0,\quad
\langle\phi_k|\phi_{l}\rangle = 0, \quad l\ne k\pm 1\quad mod \; N,\quad 
\end{eqnarray}
can  realize  the genuine  $N$-particle    interference on a multiport, governed by a genuine collective $N$-particle phase Eq.~(\ref{Rphase}) due to the permutation cycle $1\to 2 \to \ldots \to N \to 1$ and its inverse (independent from the two-particle interference parameters),   whereas  there is no  $R$-particle interference   for \mbox{$3\le R \le N-1$.}  
 
\textit{Proof.--} From the above discussion it it clear that  theorem 2 follows if there are the state vectors with the  inner products as in Eq. (\ref{E9}) and free mutual phases (thus allowing for a free $N$-particle collective phase). By Gershgorin's  circle theorem (\ref{EQ3}),  such state vectors  do exist under   the condition that \mbox{$ |\langle\phi_k|\phi_{k-1}\rangle| + |\langle\phi_k|\phi_{k+1}\rangle|  \le 1$}, for $k=1,\ldots, N$ (\textit{mod} $N$).  \mbox{Q.E.D.}

Due to the graphical representation by a polygon with the particles as the vertices and no internal edges, see fig.~\ref{F1}(b),  such single-cycle $N$-particle interference can  be termed  the ``circle-dance interference".

 Let us  further discuss an  explicit example of the circle-dance interference with  $N=4$ particles and how to detect the respective four-particle collective phase in an experiment. First, we construct the required vectors.  One can always find an  orthonormal  basis  $(|e_1\rangle, |e_2\rangle, |e_3\rangle,|e_4\rangle)$  such that the     matrix  $\Phi_{kl}  \equiv \langle e_k|\phi_l\rangle$ reads~\footnote{Since only the cyclic order $(1,2,3,4)$ of the state vectors is significant, the form of Eq. (\ref{E7}) is preserved (though in another basis) for the cyclic permutations of the state vectors.} 
\begin{eqnarray}
&&\!\!\!\! \Phi =  \left(\begin{array}{cccc}
1 & 0 & 0 & 0 \\
r_{12}e^{i\theta_{12}} & \sqrt{1-r^2_{12}} & 0 & 0 \\
0 & \frac{r_{23}}{\sqrt{1-r^2_{12}}}e^{i\theta_{23}} & \frac{\sqrt{1-r^2_{12} - r^2_{23}}}{\sqrt{1-r^2_{12}}} & 0 \\
r_{14} e^{i\theta_{14}} & - \frac{r_{12}r_{14}}{\sqrt{1-r^2_{12}}}e^{i(\theta_{14}-\theta_{12})} & \Phi_{34} & \Phi_{44}
 \end{array} \right), \nonumber\\
 && \!\!\! \Phi_{34} =  \frac{r_{34}(1-r^2_{12})e^{i\theta_{34}} + r_{12}r_{23}r_{14}e^{i(\theta_{14} - \theta_{12}-\theta_{23})}}{\left[(1-r^2_{12})(1-r^2_{12}-r^2_{23})\right]^\frac12},\nonumber\\
&& \!\!\! \Phi_{44} =  \left( \frac{1-r^2_{12} - r^2_{14}}{1-r^2_{12}} - |\Phi_{34}|^2   \right)^{\frac12},
 \label{E7}\end{eqnarray} 
 where  we use  $\Phi_{kk}$ to satisfy the normalization condition (set to be  real by the freedom of the  global phase of a quantum state) and the rest of the row elements  to satisfy the cross-state  inner  products.   Obviously,   $r_{k,k+1}$  satisfy certain conditions for the  above  construction to make sense (to satisfy the normalization of the state vectors the expressions in the square roots must be positive). While  the set   of four conditions $ r_{k,k-1} +r_{k,k+1} \le 1$  (obtained from Gershgorin's  circle theorem (\ref{EQ3})) is sufficient, an explicit analysis in this case  shows that   $ \sum_{k=1}^4 r^2_{k,k+1} \le 1$ is also sufficient (see appendix \ref{appNEW}).  For  equal  values of  $r_{k,k+1}\equiv r$ the two conditions   coincide, giving $r\le 1/2$.   

Whereas the  three-particle interference is not possible with the states in Eq.~(\ref{E7}), there is the   four-particle interference due to  the cycle $\nu_4=(1,2,3,4)$ with the  $g$-weight 
\be
g(\nu_4) =  \pm \prod_{k=1}^4 \langle\phi_k|\phi_{k+1}\rangle = \pm \exp\{ -D_{(1,2,3,4)} +i\theta_{(1,2,3,4)}\}.
\en{E8}

Besides the deterministic distinguishability of  not-the-neighbor particles on the cycle $\nu_4=(1,2,3,4)$ the  circle-dance  interference   requires the state-vectors to remain linearly independent (similar as in  three-particle case \cite{Triad2}), i.e., unambiguously distinguishable  \cite{Chefles}.  Indeed, one can verify  that  imposing linear dependence of the state vectors $|\phi_k\rangle$ (specifically  in the case of Eq. (\ref{E7}), by  setting $\Phi_{44} = 0$)  will result in a condition for   the mutual phases (making them functions of $r_{k,k+1}$).

We have stated above that any generic  multiport  can realize  the  circle-dance interference,  i.e., any multiport with no matrix element being zero. Let us consider, as an example, single photons in the symmetric four-port of Ref. \cite{ZZH} corresponding to the diamond-shaped arrangement  of four balanced beamsplitters with  one phase plate  $\varphi$ inserted into one of the internal paths, 
\be
U = \frac12 \left(\begin{array}{ccrc}
\; e^{i\varphi} & \; e^{i\varphi}  &  1 & 1\\
-e^{i\varphi} & -e^{i\varphi}  &  1 & 1\\
 -1 & \; \, 1 & -1 & 1\\
\;\, 1 & -1 & -1 & 1
\end{array}\right).
\en{U4}
In general, there are   $ \frac{N!}{R(N-R)!} $  different $R$-cycles with $N$ particles \cite{Stanley}. The symmetric group  $\S_4$  contains six 2-cycles, eight 3-cycles, and six 4-cycles, therefore, there are three permutations which are not cycles themselves, but products of disjoint 2-cycles. Since  only the neighbor  particles in the cyclic order $(1,2,3,4)$ are connected by the edges  in the corresponding graph representation,  fig. \ref{F1}(b), only the following  eight permutations, besides the trivial $I$,    contribute to the probability (divided into groups according to their cycle structure):
\begin{eqnarray}
\label{P1}
&& C_2: \; (1,2),\, (2,3),\, (3,4),\, (1,4); \nonumber\\
&& C_{2\times2}:\; (1,2)(3,4), \,(1,4)(2,3); \nonumber\\
&&C_4: \; (1,2,3,4),\, (4,3,2,1).
\end{eqnarray}
We can rearrange the summation in the  probability formula  in  Eqs. (\ref{E1}) and (\ref{E3})    to represent it as follows
\be
p_N(\bl|\bk) = \frac{1}{\m!} \sum_{\sigma\in \S_N}\per(\mathcal{U}^*(I)\circ \mathcal{U}(\sigma))\prod_{\nu\in cyc(\sigma)} g(\nu)
\en{P2}
where the symbol ``$\circ$" denotes the Hadamard (by-element) product of matrices, $\mathcal{U}_{\alpha,\beta}(\sigma) \equiv U_{k_{\sigma(\alpha)},l_\beta}$, and
\be
\per(A) = \sum_{\sigma\in\S_N} \prod_{k=1}^N A_{k,\sigma(k)}.
\en{P3}
For   $N=4$ photons   on the symmetric multiport of Eq. (\ref{U4}) with any of the phase values $\varphi\in \{ 0,\pi,\pm\pi/2\}$ and the coincidence detection, $\bl = \bk = (1,2,3,4)$,  for the cycles given in Eq. (\ref{P1}) we get
\be
\per(\mathcal{U}^*(I)\circ\, \mathcal{U}(\sigma)) = \left\{\begin{array}{cc} {3}/{32}, & \sigma \in \{I,\, C_{2\times2}\},\\
-{1}/{32}, & \sigma \in \{ C_2,\, C_4\}.
\end{array} \right.
\en{P4}
Using that  (for bosons) $g(I)=1$, $g(k,k+1) =   r^2_{k,k+1}$, $g(1,2,3,4)$ from  Eq. (\ref{E8}), and that $g(4,3,2,1) = g^*(1,2,3,4)$, we obtain from Eqs. (\ref{P2}) and (\ref{P4}) the probability for the coincidence count as follows
\begin{eqnarray}
\label{p4}
&& p_4(\bk|\bk)= \frac{1}{32}\Bigl\{ 3\left(1+ r^2_{12}r^2_{34} + r^2_{14}r^2_{23}\right) \nonumber\\
&&\quad  - \sum_{k=1}^4 r^2_{k,k+1} - 2\prod_{k=1}^4r_{k,k+1}\cos\theta_{(1,2,3,4)} \Bigr\}.
\end{eqnarray}
(For fermions, the only change in formula (\ref{p4}) is in the sign at the last two terms, due to  the minus sign at $g(k,k+1)$ and $g(1,2,3,4)$.)

Finally, we note that the   circle-dance interference is possible also   with  identical particles in mixed internal states. Indeed,    if   $\mathrm{Tr}(\rho^{(k)}\rho^{(l)})=0$ for $l\ne k\pm 1$,  then by Eq. (\ref{E11}) any  cycle $\nu$ passing  the edge $kl$ has  zero $g$-weight,  which  prevents  any  $R$-particle interference with \mbox{$3\le R\le N-1$}. 

%%%%%%%%%%%%%%%%%%%%%%%%%%%%%%%%%%%%%%%%%%%%%%%%%%%
\subsection{Four-particle  circle-dance interference with single photons having  Gaussian  spectral profiles}

Let us analyze in more detail the case of single photons having Gaussian spectral shapes and different polarizations. We consider each photon in a pure state $|\Phi_k\rangle = |\varphi_k\rangle |P_k\rangle$, where   $|P_k\rangle = \alpha_{k} |v\rangle+ \beta_{k} |h\rangle$,  with   $|v\rangle$ and $|h\rangle$ being  the polarization basis, $|\alpha_ {k}|^2 + |\beta_ {k}|^2 = 1$, and (in the frequency basis $|\omega\rangle$)
\begin{equation}
|\varphi_k\rangle =   \int d\omega \varphi_k (\omega) |\omega\rangle,
\label{vetor}
\end{equation}
with  
\begin{equation}
\varphi_k (\omega) = \left ( \frac{1}{\sqrt{ \pi} \Delta_k} \right )^{1/2} \exp{\left (- \frac{(\omega - \omega_{0k})^2}{2 \Delta_k^2} - i \tau_k (\omega - \omega_{0k})\right )}.
\label{varphi_i}
\end{equation}
 
Here we use the polarization state $|P_k\rangle$ to satisfy the necessary orthogonality conditions of theorem 2, thus we introduce an angle 
$\chi$   and  set
\begin{eqnarray}
| P_1 \rangle  =   | v \rangle,\quad | P_2    \rangle =   \cos{\chi} |v\rangle+ \sin{\chi}  |h\rangle, \nonumber\\
|P_3 \rangle = | h \rangle,  \quad  |P_4 \rangle = \sin{\chi} |v\rangle -  \cos{\chi}  |h\rangle.
\label{4vetores}
\end{eqnarray}

We also have 
\begin{equation}
\langle \varphi_k | \varphi_j \rangle = \sqrt{\frac{2 \Delta_k \Delta_j}{\Delta_k^2 + \Delta_j^2}} \hspace{1 mm} e^{-\eta_{kj}+i\xi_{kj}  } ,
\label{produtointerno}
\end{equation}
with 
\begin{eqnarray}
 \eta_{kj} &=&  \frac{1}{2} \left ( \frac{\Delta_k^2 \Delta_j^2}{\Delta_k^2 + \Delta_j^2} \right) \left[ (\tau_k - \tau_j)^2 -  \left( \frac{\omega_{0k}}{\Delta_k^2} + \frac{\omega_{0j}}{\Delta_j^2} \right)^2      \right] \nonumber \\
&+& \frac{1}{2} \left ( \frac{\omega_{0k}^2}{\Delta_k^2} + \frac{\omega_{0j}^2}{\Delta_j^2} \right ), \nonumber\\
 \xi_{kj}  &= &\left ( \frac{\Delta_k^2 \Delta_j^2}{\Delta_k^2 + \Delta_j^2} \right )\left( \frac{\omega_{0k}}{\Delta_k^2} + \frac{\omega_{0j}}{\Delta_j^2} \right) (\tau_j - \tau_k)\nonumber\\
  & +& (\tau_k \omega_{0k} - \tau_j \omega_{0j}).
\label{coeficiente1}
\end{eqnarray} 
From the definition    $\langle \Phi_k | \Phi_j \rangle=r_{kj} e^{ i\theta_{kj}}$,  we get
\begin{eqnarray}
r_{12} &=& \cos{\chi} \sqrt{\frac{2 \Delta_1 \Delta_2}{\Delta^2_1 + \Delta^2_2}} e^{\eta_{12}}, \quad  \theta_{12} =  \xi_{12},  \nonumber\\
r_{23} &=& \sin{\chi} \sqrt{\frac{2 \Delta_2 \Delta_3}{\Delta^2_2 + \Delta^2_3}} e^{\eta_{23}} , \quad   \theta_{23} =  \xi_{23}, \nonumber\\
r_{34} &=& \cos{\chi} \sqrt{\frac{2 \Delta_3 \Delta_4}{\Delta^2_3 + \Delta^2_4}} e^{\eta_{34}}, \quad    \theta_{34} = \pi + \xi_{34},   \nonumber\\
r_{41} &=& \sin{\chi} \sqrt{\frac{2 \Delta_4 \Delta_1}{\Delta^2_4 + \Delta^2_1}} e^{\eta_{14}},  \quad  \theta_{41} =  \xi_{41}.
\label{r_kj}
\end{eqnarray}

For $\Delta_k\ne \Delta_l$ the analysis   becomes quite involved, thus we consider below   all spectral width  being the same, $\Delta_k = \Delta$. We have in this case
\be
\eta_{k,k+1} =  \frac{(\omega_{0k}-\omega_{0,k+1})^2}{4\Delta^2}+\frac{\Delta^2}{4}(\tau_k - \tau_{k+1})^2
\en{eta_kj}
and 
\be
 \xi_{k,k+1}=\frac{1}{2} (\tau_k + \tau_{k+1})(\omega_{0k}-\omega_{0,k+1}).
\en{xi_kj}
From Eqs. (\ref{Rphase}),  (\ref{r_kj})   and (\ref{xi_kj})  the four-particle collective phase  becomes
\begin{eqnarray} 
\label{4Phase}
\theta_{(1,2,3,4)} &=&  \pi+  \frac12 \Bigl[ (\omega_{04}-\omega_{02})(\tau_1-\tau_3) \nonumber\\
&&+ (\omega_{03}-\omega_{01})(\tau_4-\tau_2)\Bigr].
 \end{eqnarray}
  Eqs. (\ref{eta_kj}) and (\ref{4Phase})  involve the differences of  four frequencies and four arrival times,   hence they contain  only six real parameters additionally to the polarization angle $\chi$ of Eq. (\ref{4vetores}).  One can therefore arrange  for the  two-particle  interference  parameters $r_{kj}$  (\ref{r_kj}) to remain fixed, which takes up only four free parameters  from the total seven, thus leaving   the four-particle collective  phase (\ref{4Phase})  to vary on  a three-parameter manifold.  Therefore, the four-particle circle dance interference can be  observed with the  photons in Gaussian spectral profiles when one can arrange for  variable central frequencies of the spectral states and  the photon arrival times. 
 
 The above  example serves only for illustration, since it may be experimentally challenging. There are other possible ways to arrange for the circle dance interference, one such  way is reported elsewhere \cite{Cleo}.

%%%%%%%%%%%%%%%%%%%%%%%%%%%%%%%%%%%%%%%%%%%%%%%
\section{The circle-dance interference and multiparticle correlations}
\label{sec4}

What is the reason behind the fact that the collective phase of the circle-dance interference of $N$ particles cannot be detected in an experiment with less than $N$ particles (irrespectively if some particles are simply not sent to a multiport,  there are particle  losses,   or one bins together the output configurations containing a given configuration with  less than $N$ particles)? Here we show that such a phase is  a signature of the  genuine $N$th-order quantum correlations between them, i.e., the correlations not reducible to the lower-order ones. 

We will use the fact that detection  of $R$ photons  is related to the $R$-th order correlation function  \cite{Glauber}  of the quantum field. This relation  extends also  to our case, when   the internal states of identical particles are not resolved  (which corresponds to a generalized measurement, mathematically expressed by summing up the probabilities with the resolved internal states). Therefore,  we can adopt as the  unnormalized $R$th order correlation function $Q_R(\bl)$  in $R$ output ports $\bl^\prime=(l_1,\ldots,l_R)$  the following expression
\be
Q_R(\bl^\prime) = \left\langle \sum_{\mathbf{j}}\left(\prod_{\alpha=1}^R \hat{b}^\dag_{l_\alpha, j_\alpha} \right)\left(\prod_{\alpha=1}^R\hat{b}_{l_\alpha, j_\alpha} \right)\right\rangle,
\en{CF}
where we  sum over the basis states in  $\mathcal{H}^{\otimes R}$, $\hat{b}^\dag_{l,j}$  creates a particle in output port $l$ of a multiport and an internal state $j$,  and the average is taken with  the input state. For  $N$ particles in  input $\bk$,  the corresponding  function  $Q_R(\bl^\prime|\bk)$, as the probability $p_N(\bl^\prime|\bk)$  in Eq.  (\ref{E10}), depends only on the $d$-cycles with $d\le R$, since the two are proportional   (see  appendix \ref{appE})
\be
Q_R(\bl^\prime|\bk)  = \m^\prime! \left({N \atop R}\right)p_N(\bl^\prime|\bk) =  \m^\prime! \sum_{ \bk^{\prime}\subset \bk} p_R(\bl^{\prime}|\bk^{\prime}).
\en{Id}
 By   Eq. (\ref{Id})  all $Q_R(\bl^\prime|\bk)$ for $R\le N-1$ are independent of the $N$-particle collective phase of Eq. (\ref{Rphase}). 
 
 Note that $Q_R(\bl^\prime|\bk)$ contains both  quantum and classical correlations,  in the classical case (distinguishable particles in orthogonal internal states) it  is proportional to the respective   classical probability. The   classical correlation function is therefore independent of the collective phases (since   particles in the orthogonal internal states do not have collective phases). 
 
Eq. (\ref{Id}) together with the results of the previous sections  means that some  popular  criteria for distinguishing quantum and classical behavior of identical particles in unitary linear multiports  as in Ref.  \cite{Wal} and   the recently introduced nonclassicality criteria for interference  in multiport interferomentry \cite{Nonclass}, based on the  second-order correlation,   will not be able to detect  quantum $R$-particle phases, for all $R\ge 3$, since they are related to higher than the second-order correlations. Therefore, though such criteria may detect some quantum behavior at a multiport output,  they are far from being sufficient for this purpose.

\subsection{Discrimination  of  internal states  and  multiparticle  interference}
\label{sec4A}
Up to now we have assumed that particle detection at the output of a linear multiport  does not  resolve the  internal states of particles. Let us now  discuss  an internal state  resolving detection (see appendix \ref{appA} for the mathematical details of state resolving  detection). 

If   a detector  resolves the  internal state of at least one particle, the   $N$-particle  interference   does not contribute to the respective probability. Let us illustrate this using  the   circle-dance $N$-particle interference with identical particles in pure internal states, which   requires the internal states to be linearly independent, i.e., unambiguously distinguishable by the scheme of Ref. \cite{Chefles}.     The   unambiguous discrimination scheme of Ref. \cite{Chefles} runs as follows.  The   state $|\phi_k\rangle$ is identified with some probability $p_k$,  the corresponding measurement  operator being  $\Pi_k = p_k|\phi^{(\perp)}_k\rangle\langle \phi^{(\perp)}_k|$,  where $\langle\phi^{(\perp)}_k|\phi_l\rangle = \delta_{kl}$, whereas  an inconclusive result corresponds to $\Pi_0 = \openone- \sum_{k=1}^N \Pi_k$. In the graph representation, see fig. \ref{F1}(b), even a single such an internal state detection  with a  conclusive result  implies that vertex $k$     has  no edges (since a particular input particle is detected at an  output port, it  does not participate in the permutations of particles in the quantum amplitudes, as discussed in section \ref{sec2}). Broken edge in the cycle of fig. \ref{F1}(b) means that  the  $N$-particle  interference is not observed (neither the  $(R\ge 3)$-particle interference in this case). The inconclusive result, on the other hand, does not affect the terms in the probability  coming from the $(R\ge 2)$-cycles, due to
$\langle \phi_k|\Pi_0|\phi_l\rangle = \langle \phi_k|\phi_l\rangle$ for $k\ne l$, (the edges of the respective graph retain  their weights), while   simultaneously attenuating the permutations with fixed points, since for each fixed point  we have in this case  $\langle \phi_k|\Pi_0|\phi_k\rangle  = 1-p_k$, instead of $g(k)=1$ in the state non-resolving detection. Since  in the case of  the circle-dance interference   there are  just the two-cycles $(k,k+1)$ and the full $N$-cycle,  such a generalized detection  attenuates the two-particle interference (the permutations having fixed points). Thus the unambiguous state discrimination would separate the outcomes with either  destructed  or relatively enhanced  circle-dance interference. 

The above discussion indicates that the  internal state discrimination, even if partial   (e.g., the resolution of the arrival times of single photons),   can strongly affect the multiparticle  interference.    Whether one is technically able to perform generalized measurements on the internal states of identical particles, as  required in the above  unambiguous state discrimination  scheme,   is another problem  that depends on the particular physical setup, the type of identical particles used and the degrees of freedom that serve as the Hilbert space of  the internal states.

\section{Conclusion}
\label{sec5}

 We have provided a general framework which allows to study   the  complex relation between particle distinguishability and higher-order interference effects  of  independent identical bosons or  fermions on a linear multiport.  We have introduced  the collective geometric  phases of identical particles, which govern multiparticle interferences,  and related them to the  higher-order quantum correlations acquired between independent particles via   propagation in a linear  multiport.      We also have opened the discussion on  exact relation between the  state-discrimination distinguishability and the   distinguishability in the multiparticle interference, by showing that  the   genuine $N$-particle interference for $N\ge 4$ independent particles in pure internal states   requires   each particle  to be  unambiguously distinguishable from all  others and deterministically distinguishable from all but two. We show, for instance, that   the unambiguous internal state discrimination    is detrimental to  the interference. However,  the latter gets an enhanced visibility,  if  the measurement result   is inconclusive. 
 
Throughout the work we have seen an interesting  connection of the partial distinguishability  theory to the theory of  weighted graphs. We show,  for instance,  how the usual concept of a weighted directed graph with $N$ vertices appears in the interference of $N$ identical particles  in pure internal states, where the weights are defined by the inner products of the internal states. Though   one could  also obtain all our  results by using only the combinatorics with permutations, this connection by itself is very interesting. Indeed,  the weighted graph theory is intimately linked with one of most studied  computationally hard problems, namely,  the traveling salesman problem  \cite{TSP}.   Such a connection and what it has to say about the computational complexity of partially distinguishable identical particles  is  worth exploring in the future publications.

 \medskip
\section{Acknowledgements}  
V.S.  was supported by the National Council for Scientific and Technological Development (CNPq) of Brazil,  grant  304129/2015-1, and by  the S{\~a}o Paulo Research Foundation   (FAPESP), grant 2015/23296-8.  M. E. O. was supported by  FAPESP, grant number 2017/06774-9.

\medskip
%\widetext

\appendix
\section{ Brief derivation  of Eqs. (\ref{E1}) and (\ref{E3})}
\label{appA}

 Here we provide  brief derivation of output  probability formula which reduces to  Eqs. (\ref{E1}) and (\ref{E3})  for particle counting nonresolving the internal states.  We follow Ref.  \cite{PartDist} and the Supplemental Material to Refs. \cite{BB,GL}. Below we  consider simultaneously  bosons and fermions using the fact that the respective probability  is  identical for the same $J$-function (the internal states for the same $J$-function are obviously  not the same).

The general possible  mixed state of   $N$ partially-distinguishable    particles impinging at $\bk=(k_1,\ldots,k_N)$ input ports   of  a unitary $M$-port $U$, corresponding to the occupations $\mathbf{n} = (n_1,\ldots,n_M)$,  can always be cast as follows 
 \be
{\rho}(\mathbf{n}) = \sum_{i}p_i|\Psi_i\rangle\langle\Psi_i|,
 \quad |\Psi_i \rangle = \frac{1}{\sqrt{\mathbf{n}!}}\sum_{\mathbf{j}}C^{(i)}_{\mathbf{j}}\prod_{\alpha=1}^N \hat{a}^\dag_{k_\alpha,j_\alpha}|0\rangle,
\en{A0}
where $\hat{a}^\dag_{k,j}$ is  the  creation  operator of a boson (fermion)  in port $k$ and an internal  basis state $|j\rangle \in\mathcal{H}$, $\mathbf{n}!  = n_1!\ldots n_M!$, $\mathbf{j} = (j_1,\ldots,j_N)$,  and $p_i\ge 0$ with  $\sum_i p_i =1$. The permutation symmetry (anti-symmetry) of the  operators for bosons (fermions)  allows one to choose  the expansion coefficients $C^{(i)}_{\mathbf{j}}$    to be symmetric     (anti-symmetric) with respect to the symmetry subgroup $\S_{\mathbf{n}}\equiv \mathcal{S}_{n_1}\otimes \ldots \otimes \mathcal{S}_{n_M}$ of the symmetric group $\mathcal{S}_N$, where $\mathcal{S}_{n_k}$ corresponds to the  permutations of the  internal states of the particles    in   input port $k$ between themselves.   The  symmetric (anti-symmetric) coefficients are normalized by  $\sum_{\mathbf{j}} |C^{(i)}_{\mathbf{j}}|^2 =1$.

A  linear multiport $U$ performs  the unitary transformation $\hat{a}^\dag_{k,j} = \sum_{l=1}^M U_{kl}\hat{b}^\dag_{l,j}$.  Consider a particle  counting at the multiport output capable of not only the  particle-number, but also the   internal state resolution. First, let us  assume that the internal states are resolved by the projective measurement in some orthogonal basis (let it be the same as in Eq. (\ref{A0})).
The  probability of  the output  $(l_1,j_1),\ldots,(l_N,j_N)$  in the $b$-basis, with   the corresponding particle counts  $s_{l,j}\ge 1$,   reads
\be
p_N(\bl,\mathbf{j}|\bk) = \mathrm{Tr}( {\rho}(\mathbf{n}) \mathcal{D}(\mathbf{s})).
\en{A01}
where
\be
 \mathcal{D}(\mathbf{s}) =   \frac{1}{\mathbf{s}!} \left[\prod_{\alpha=1}^N \hat{b}^\dag_{l_\alpha,j_\alpha}\right]|0\rangle\langle0|\left[\prod_{\alpha=1}^N \hat{b}_{l_\alpha,j_\alpha}\right].
\en{A02}
Using the  following  identity
\begin{eqnarray}
\label{A03}
&& \langle0|\left[\prod_{\alpha=1}^N \hat{b}_{l_\alpha,j_\alpha}\right]\left[\prod_{\alpha=1}^N \hat{b}^\dag_{l^\prime_\alpha,j^\prime_\alpha}\right]|0\rangle\nonumber\\
&& = \sum_{\sigma\in\S_N}\vare(\sigma)\prod_{\alpha=1}^N  \delta_{l^\prime_\alpha,l_{\sigma(\alpha)}} \delta_{j^\prime_\alpha,j_{\sigma(\alpha)}},
\end{eqnarray}
with   $\vare(\sigma) = 1$ for bosons and $\vare(\sigma) = \mathrm{sgn}(\sigma)$   for fermions,    one obtains
\be
p_N(\bl,\mathbf{j}|\bk)  = \frac{1}{\mathbf{s}!\mathbf{n}!} \sum_{\tau,\sigma\in \S_N}J(\mathbf{j};\tau^{-1},\sigma)
  \prod_{a=1}^N U^*_{k_\alpha,l_{\tau(\alpha)}}U_{k_\alpha,l_{\sigma(\alpha)}},
\en{A04}
where the   $J$-function reads
\be
J(\mathbf{j};\tau,\sigma) = \vare(\sigma\tau)\mathrm{Tr}\left( \Pi_{j_1} \otimes\ldots\otimes \Pi_{j_N} P_\sigma \rho^{(int)} P_\tau \right)
\en{GenJ}
 with  $\Pi_j = |j\rangle\langle j|$ and the  internal state  $\rho^{(int)}$ given as
\be
\rho ^{(int)}= \sum_i p_i |\psi_i\rangle\langle\psi_i|,\quad |\psi_i\rangle \equiv \sum_{\mathbf{j}} C^{(i)}_{\mathbf{j}}  \prod_{\alpha=1}^N{\!}^{\otimes}|j_\alpha\rangle.
\en{rho_int}
Note the symmetry property of the  $J$-function.  The permutation symmetry (anti-symmetry) of the internal  state (\ref{rho_int})  for bosons (fermions), i.e., $P_\pi \rho^{(int)} = \rho^{(int)} P_\pi = \vare(\pi)\rho^{(int)}$ for any $\pi\in \S_{\mathbf{n}}$, implies that
\be
J(\mathbf{j}; \tau,\sigma\pi) = J(\mathbf{j}; \pi\tau,\sigma)=  J(\mathbf{j}; \tau,\sigma), \quad  \forall \pi \in \S_{\mathbf{n}}.
\en{Jsym}

For independent particles $\rho^{(int)} = \rho^{(1)}\otimes\ldots\otimes\rho^{(N)}$ the expression for the $J$-function in Eq. (\ref{GenJ}) can be simplified. Using that
\be
P^\dag_{\sigma}\left(\prod_{\alpha=1}^N \Pi_\alpha \right) P_\sigma = \prod_{\alpha=1}^N \Pi_{\sigma(\alpha)}
\en{Ident}
 we get
\begin{eqnarray}
\label{Jind}
&& J =   \vare(\sigma\tau)\mathrm{Tr}\left( P^\dag_{\sigma^{-1}\tau} \Pi_{j_{\sigma(1)}}\rho^{(1)} \otimes\ldots\otimes \Pi_{j_{\sigma(N)}} \rho^{(N)} \right)\nonumber\\
 && = \prod_{\nu} (\pm1)^{|\nu|-1}\mathrm{Tr}\left(\Pi_{j_{\sigma(\alpha_1)}}\rho^{(\alpha_1)} \ldots  \Pi_{j_{\sigma(\alpha_{|\nu|})}}\rho^{(\alpha_{|\nu|})} \right),\nonumber\\
\end{eqnarray}
where $\nu=(\alpha_1,\ldots,\alpha_{|\nu|})$ is from  the disjoint  cycle decomposition of  $\sigma^{-1}\tau$,  we have used that $ \vare(\nu) = (\pm1)^{|\nu|-1}$ \cite{Stanley} and  the identity 
\begin{eqnarray}
\label{A05}
&& \mathrm{Tr}(P^\dag_\sigma A^{(1)}\otimes \ldots \otimes A^{(N)}) \nonumber\\
&& =  \prod_{\nu\in cyc(\sigma)}\mathrm{Tr}(A^{(\alpha_1)}\otimes \ldots \otimes A^{(\alpha_{|\nu|})}).
\end{eqnarray}

Though we have considered above  the specific projective measurement,   $\Pi_j = |j\rangle\langle j|$, by the linearity property Eqs. (\ref{GenJ})-(\ref{Jind})  apply also for arbitrary POVM elements $\Pi_j$, where $\sum_j \Pi_j = \openone$.

For the particle counting not resolving their internal states, which is  obtained  by    summation of the probabilities with all possible   $\Pi_j$ elements   in \
Eq. (\ref{A04}) for  a fixed set of the output ports $\bl=(l_1,\ldots,l_N)$,  the  probability  reads
\begin{eqnarray}
\label{pNONR}
&& p_N(\bl|\bk) = \sum_{\mathbf{j}}p_N(\bl,\mathbf{j}|\bk)  \\
&& =  \frac{1}{\mathbf{m}!\mathbf{n}!} \sum_{\tau,\sigma\in \S_N} J(\tau^{-1}\sigma)\prod_{a=1}^N U^*_{k_\alpha,l_{\tau(\alpha)}}U_{k_\alpha,l_{\sigma(\alpha)}},\nonumber
\end{eqnarray}
with  $\m = (m_1,\ldots,m_M)$ being the corresponding  occupations  and
\be
J(\sigma) =    \vare(\sigma) \mathrm{Tr}\bigl(  P_{\sigma}\rho^{(int)}\bigr).
\en{A013}
For  independent particles and single particle per input port, from Eqs. (\ref{A05}), (\ref{pNONR}) and (\ref{A013})  we obtain   Eqs. (\ref{E1})-(\ref{E3}) of section \ref{sec2}.

%%%%%%%%%%%%%%%%%%%%%%%%%%%%%%%%%%%%%%
\section{The probability to detect $R$ out of $N$ particles}
\label{appD}

Here we give a direct mathematical derivation of the probability formula in Eq. (\ref{E10}), in particular, we prove the fact that the marginal probability of detecting $R< N$ particles from $N$ input ones depends \textit{only} on the cycles with length not exceeding $R$.

First of all, since  identical particles bear no labels, we have (see also the note in  section \ref{sec2}) 
\be
p_N(\sigma(\bl)|\bk)= p_N(\bl|\sigma(\bk)) = p_N(\bl|\bk), \quad \sigma \in \S_N,
\en{B2}
where $\sigma(\bl) = (l_{\sigma(1)},\ldots,l_{\sigma(N)})$ (and similar for $\bk$). For definiteness, let us assume  that $\bk$ are    ordered,   $k_\alpha\le k_{\alpha+1}$. 

Consider now  a symmetric function in $N$ variables $f(\bl)$. Denote by $m_k$ the number of  occurrences of $l_\alpha = k$ and set $\m = (m_1,\ldots,m_M)$. Partitioning $\bl = (\bl^{\prime},\bl^{\prime\prime})$ such that $\bl^{\prime} = (l_1,\ldots,l_{R})$ (and, correspondingly, $\m = (\m^\prime,\m^{\prime\prime})$)   we get
\begin{eqnarray}
\label{B3}
&&\sum_{\m} f(\bl) = \sum_{\bl} \frac{\m!}{N!} f(\bl) = \frac{1}{N!}\sum_{\bl^{\prime}}\sum_{\bl^{\prime\prime}} \m! f(\bl)\nonumber\\
&&= \left({N \atop R}\right)^{-1}\sum_{\m^{\prime}}\sum_{\m^{\prime\prime}} \frac{\m!}{\m^{\prime}!\m^{\prime\prime}!} f(\bl),
\end{eqnarray}
where the summation runs  over  $|\m| \equiv m_1+\ldots m_M = N$, $|\m^{\prime}|=R$, and $|\m^{\prime\prime}|=N-R$. Eq. (\ref{B3}) applied to the probability $p_N(\bl|\bk)$ supplies the expression for  the probability to detect $R$ particles out of $N$, corresponding to  occupations $\m^{\prime}$
\be
p_N(\bl^{\prime}|\bk) \equiv \left({N \atop R}\right)^{-1}\sum_{\m^{\prime\prime}} \frac{\m!}{\m^\prime!\m^{\prime\prime}!}p_N(\bl|\bk) ,
\en{B4}
where $\bl$ is partitioned as above. 

Since the probability $p_N(\bl|\bk)$ is symmetric in both $\bk$ and $\bl$, we can rearrange $\bl$ such  that $\bl^\prime = (l_1,\ldots,l_R)$. Now we plug the general formula Eq. (\ref{pNONR}) (in an equivalent form)  into Eq. (\ref{B4}):
\begin{eqnarray}
\label{B5}
&& p_N(\bl^{\prime}|\bk) = \frac{R!}{N!\m^{\prime}!\n!} \sum_{\sigma_{1,2}\in \S_N} J(\sigma_1\sigma^{-1}_2)\nonumber\\
&& \times \sum_{\bl^{\prime\prime}}\prod_{\alpha=1}^NU^*_{k_{\sigma_1(\alpha)},l_\alpha}
U_{k_{\sigma_2(\alpha)},l_\alpha} \nonumber\\
&& = \frac{R!}{N!\m^{\prime}!\n!} \sum_{\sigma_{1,2}\in \S_N} J(\sigma_1\sigma^{-1}_2)\left[\prod_{\alpha=1}^RU^*_{k_{\sigma_1(\alpha)},l_\alpha}
U_{k_{\sigma_2(\alpha)},l_\alpha} \right]\nonumber\\
&& \times \prod_{\alpha=R+1}^N\delta_{k_{\sigma_1(\alpha)},k_{\sigma_2(\alpha)}},
\end{eqnarray}
where we have used the unitarity of the multiport matrix $U$. Let us decompose the permutations $\sigma_{1,2}$  as follows
\be
\sigma_i = (\sigma_i^\prime\otimes\sigma^{\prime\prime}_i)\tau_i,\quad \sigma_i^{\prime}\in \S_R,\quad \sigma_i^{\prime\prime}\in \S_{N-R},
\en{B6}
i.e., we define $\tau_i\in \S_N/(\S_R\S_{N-R})$, the permutation choosing two subsets of $R$ and $N-R$ indices from $1\le \alpha \le N$ without changing the proper order of the indices in each  subset.   Then $\tau_i (\bk)\equiv (k_{\tau_i(1)},\ldots,k_{\tau_i(N)}) =
(\bk^{\prime(i)},\bk^{\prime\prime(i)})$ with  $k^{\prime}_\alpha \le k^{\prime}_{\alpha+1}$ and $k^{\prime\prime}_\alpha \le k^{\prime\prime}_{\alpha+1}$ (by our choice of the order of $k_\alpha$). We have
\begin{eqnarray}
\label{B7}
&&\prod_{\alpha=R+1}^N\delta_{k_{\sigma_1(\alpha)},k_{\sigma_2(\alpha)}} = \delta_{\bk^{\prime(1)},\bk^{\prime(2)}}\prod_{\alpha=1}^{N-R}
\delta_{k^{\prime\prime}_{\sigma^{\prime\prime}_1(\alpha)},k^{\prime\prime}_{\sigma^{\prime\prime}_2(\alpha)}}\nonumber\\
&&\qquad = \delta_{\bk^{\prime(1)},\bk^{\prime(2)}}\sum_{\pi\in\S_{\n^{\prime\prime}}}\delta_{\sigma^{\prime\prime}_1(\sigma^{\prime\prime}_2)^{-1},\pi}
\end{eqnarray}
where $\S_{\n^{\prime\prime}}\equiv \S_{n^{\prime\prime}_{1}}\otimes\ldots\otimes\S_{n^{\prime\prime}_M}$ is the symmetry group of the input configuration $\bk^{\prime\prime}$ with the occupations $\n^{\prime\prime}$.
For $\pi\in \S_\n$ and an arbitrary $\sigma\in \S_N$ we have from Eq. (\ref{Jsym})
\be
J(\sigma\pi) = J(\pi\sigma) = J(\sigma).
\en{B8}
In our case
\be
\sigma_2\sigma^{-1}_1 = \sigma^\prime_2(\sigma^{\prime}_1)^{-1}\otimes \pi^{\prime\prime} = (\sigma^\prime_2(\sigma^{\prime}_1)^{-1}\otimes I)\pi ,
\en{B9}
with $\pi \equiv I\otimes\pi^{\prime\prime} \in \S_\n$. Now note  the  following  identity,  valid for any symmetric function $f(\bk^\prime)$,
\begin{eqnarray}
\label{id2use}
&& \sum_{\sigma_{1,2}} \delta_{\bk^{\prime(1)},\bk^{\prime(2)}}\sum_{\pi\in\S_{\n^{\prime\prime}}}\delta_{\sigma^{\prime\prime}_1(\sigma^{\prime\prime}_2)^{-1},\pi}f(\bk^\prime) \nonumber\\
&&\qquad= (N-R)! \sum_{\bk^\prime \subset \bk}\frac{\n!}{\n^\prime!}f(\bk^\prime),
\end{eqnarray}
where the sum over permutations is reduced to that over values $\bk^\prime$, we have used that  exactly $\frac{\n!}{\n^\prime!\n^{\prime\prime}!}$ of $\tau_{2}$ give the same set of values  $\bk^{\prime(2)}$  as $\bk^{\prime(1)}$, the number of permutations $\pi\in \S_{\n^{\prime\prime}}$  and that one $\sigma^{\prime\prime}_i\in \S_{N-R}$  can be chosen arbitrary. By using the identity from Eq. (\ref{id2use}),  the fact that the summation over $\tau_1$ is equivalent to that over  the  ordered subsets $\bk^\prime\subset \bk$ (or over all choices of  $R$ particles   from  $N$), and that there are exactly $\n^{\prime\prime}!$ permutations     in $\S_{\n^{\prime\prime}}$, we obtain from Eqs. (\ref{B5})-(\ref{B9}):
\be
p_N(\bl^{\prime}|\bk) =\left({N \atop R}\right)^{-1}\sum_{\bk^\prime\subset \bk}p_R(\bl^\prime|\bk^\prime),
\en{B10}
where we have introduced the probability  of the particles from input $\bk^\prime$ to be detected at the output ports $\bl^\prime$
\begin{eqnarray}
p_R(\bl^\prime|\bk^\prime)&=&\frac{1}{\m^{\prime}!\n^\prime!} \sum_{\sigma_{1,2}\in\S_R}J_{\n^\prime}(\sigma_1\sigma^{-1}_2)\nonumber\\
&&\times\prod_{\alpha=1}^R U^*_{k_{\sigma_1(\alpha)},l_\alpha}U_{k_{\sigma_2(\alpha)},l_\alpha}.
\end{eqnarray}
Here  $J_{\n^\prime}$-function  is defined as in Eq. (\ref{E3}) of section \ref{sec2}  for the internal states of the particles  labelled by $\bk^\prime$.  

Since  the probability in Eq. (\ref{B10})  is given by averaging the respective probabilities $p_R(\bl^\prime|\bk^\prime)$ with input of $R$ particles (selected arbitrarily from $N$), therefore it depends only on the cycles of length $d\le R$.

%%%%%%%%%%%%%%%%%%%%%%%%%%%%%%%%%%%%
\section{Bound on the  $g$-weight}
\label{appB}

The bound is valid for arbitrary positive semi-definite Hermitian operators $A_1,\ldots,A_n$, i.e.,
\be
|\mathrm{Tr}(A_1\ldots A_n)|^2 \le \prod_{k=1}^n \mathrm{Tr}(A_kA_{k+1}),
\en{C1}
where $k$ is $\textit{mod}\; n$. Note that $\mathrm{Tr}(A_kA_{k+1})$ is real and positive.  Eq. (\ref{C1}) can be proven by using the Cauchy-Schwartz inequality
for the Hilbert-Schmidt (a.k.a.  Frobenius)  norm, i.e., for any two operators $A$ and $B$ having  finite Frobenius norm
\be
|\mathrm{Tr}(A^\dag B)| \le \sqrt{\mathrm{Tr}(A^\dag A)}\sqrt{\mathrm{Tr}(B^\dag B)}\equiv ||A||\cdot||B||.
\en{C2}
Now let us introduce the positive semi-definite Hermitian operator  $A^{\frac12}_k$, i.e., $A^{\frac12}_kA^{\frac12}_k = A_k$.  Rearranging  the factors by placing  one $A^{\frac12}_1$ to the right, we have
\begin{eqnarray}
\label{C3}
 &&|\mathrm{Tr}(A_1\ldots A_n)| = |\mathrm{Tr}\{ (A^{\frac12}_1A^{\frac12}_2) (A^{\frac12}_2A^{\frac12}_3)\ldots ( A^{\frac12}_n A^{\frac12}_1)\}| \nonumber\\
&&  \le ||A^{\frac12}_1A^{\frac12}_2|| \ldots ||A^{\frac12}_nA^{\frac12}_1|| = \prod_{k=1}^n \sqrt{\mathrm{Tr}(A_kA_{k+1})},
\end{eqnarray}
where we have used Eq. (\ref{C2}), the submultiplicativity of the Frobenius norm $ ||AB|| \le ||A||\cdot ||B||$ and the effective commutativity  when evaluating  the trace of a product of two operators.
%%%%%%%%%%%%%%%%%%%%
\section{Roots of the matrix $H$ for $N=4$}
\label{appNEW}

In this case we have for the characteristic equation for $H_{kl} = \langle\phi_k|\phi_l\rangle= r_{kl}e^{i\theta_{kl}}$:
\be 
\mathrm{det}(H - \lambda I) = (\lambda-1)^4 - (\lambda-1)^2a_2 + a_0,
 \en{N1}
 with $a_2 = r^2_{12}+r^2_{23} +r^2_{34} +r^2_{14}$ and 
 \be
 a_0 = | r_{12}r_{34}e^{i(\theta_{12}+\theta_{34})} - r_{14}r_{23}e^{i(\theta_{14}+\theta_{32})}|^2.
 \en{N2}
 We need to require that $\lambda \ge 0$ for the matrix $H$ to be positive semidefinite Hermitian matrix. For the roots  we get  
 \be
 (\lambda-1)^2 = \frac12\left( a_2 \pm \sqrt{a_2- 4a_0}\,\right).
 \en{N3}
Now, using that $a^2_2 \ge 4a_0$ (the matrix $H$ is Hermitian, therefore the  roots  are real)  we get from Eq. (\ref{N3})  that $a_2\le 1$ is   a sufficient condition for the matrix $H$ to be positive semidefinite. 
 
%%%%%%%%%%%%%%%%%%%%%%%%%%%%%%%%%%%%%%
\section{The expression for the correlation function}
\label{appE}

Note that following expression for  the projector on the symmetric (anti-symmetric) subspace of $N-R$ particles (when one discards the information on the internal states in $\mathcal{H}$)
\be
S_{N-R} = \sum_{\m^{\prime\prime}} \frac{1}{\m^{\prime\prime}!} \sum_{\mathbf{j}} \left(\prod_{\alpha=1}^{N-R} \hat{b}^\dag_{l^{\prime\prime}_\alpha, j_\alpha} \right)|0\rangle\langle 0| \left(\prod_{\alpha=1}^{N-R} \hat{b}_{l^{\prime\prime}_\alpha, j_\alpha} \right)
\en{D1}
where $\m^{\prime\prime}$ is the occupations  of  $N-R$ particles in the output ports $\bl^{\prime\prime}=(l^{\prime\prime}_1,\ldots, l^{\prime\prime}_{N-R})$ (in the partition $\bl = (\bl^\prime,\bl^{\prime\prime})$). Using the fact that    the correlation function is computed for an $N$-particle state $\rho(\n)$ (symmetric or anti-symmetric under particle permutations, respectively, for bosons and fermions)  by inserting  the projector of Eq.~(\ref{D1}) into the expression in   Eq.~(\ref{CF}) we get
\begin{eqnarray}
\label{D2}
&&Q_R(\bl^\prime|\bk) =\left\langle \sum_{\mathbf{j}}\left(\prod_{\alpha=1}^R \hat{b}^\dag_{l^\prime_\alpha, j_\alpha} \right)S_{N-R}\left(\prod_{\alpha=1}^R \hat{b}_{l^\prime_\alpha, j_\alpha} \right)\right\rangle\nonumber\\
&&= \left\langle \sum_{\m^{\prime\prime}}\frac{1}{\m^{\prime\prime}!} \sum_{\mathbf{j}} \left(\prod_{\alpha=1}^N \hat{b}^\dag_{l_\alpha, j_\alpha} \right)|0\rangle\langle 0| \left(\prod_{\alpha=1}^N \hat{b}_{l_\alpha, j_\alpha} \right)\right\rangle\nonumber\\
&&=\sum_{\m^{\prime\prime}}\frac{\m!}{\m^{\prime\prime}!} p_N(\bl|\bk),
\end{eqnarray}
since    starting from the second line in Eq. (\ref{D2}) the  evaluation of the  average is reduced to   the   steps  similar to those in the derivation of Eq. (\ref{pNONR}) in appendix \ref{appA}. Comparing  Eqs. (\ref{D2}) and  (\ref{B4}) we obtain
\be
Q_R(\bl^\prime|\bk) = \m^\prime! \left({N \atop R}\right)p_N(\bl^\prime|\bk),
\en{D3}
i.e., the first formula  in Eq. (\ref{Id}), where the second formula follows from Eq. (\ref{B10}).
 
%%%%%%%%%%%%%%%%%%%%%%%%%%%%%%%%%%%%%%%%%%%%%%%%%%%%

\end{document}